\documentclass[conference,10 pt,letterpaper]{ieeeconf}
\IEEEoverridecommandlockouts 
\usepackage{amssymb}
\usepackage{amsmath}
\usepackage{bm}

\title{\LARGE \bf A generalized spatiotemporal covariance model \\ for stationary
  background in analysis of MEG data} 

\author{Plis S.M., {\it Student Member, IEEE}, Schmidt D.M., Jun S.C., {\it Member, IEEE}, Ranken D.M.%
\thanks{This work was supported by NIH  grant 2 R01 EB000310-05 and the Mental
Illness  and Neuroscience  Discovery (MIND)  Institute.}
\thanks{Plis S.M. is with Computer Science Department,
University of New Mexico, Albuquerque, NM 87131, USA
and with Biological \& Quantum Physics Group, Los Alamos National Laboratory,  
Los Alamos,NM 87545, USA {\tt\small pliz@cs.unm.edu}}
\thanks{Schmidt D.M. is with Biological \& Quantum Physics Group, 
Los Alamos National Laboratory,  Los Alamos,NM 87545, USA}
\thanks{Jun S.C. is with Biological \& Quantum Physics Group, 
Los Alamos National Laboratory,  Los Alamos,NM 87545, USA}
\thanks{Ranken D.M. is with Biological \& Quantum Physics Group, 
Los Alamos National Laboratory,  Los Alamos,NM 87545, USA}
}

\begin{document}
\maketitle

\begin{abstract}
  Using a noise  covariance model based on a  single Kronecker product
  of spatial and temporal covariance in the spatiotemporal analysis of
  MEG data was demonstrated to provide improvement in the results over
  that of the commonly used  diagonal noise covariance model.  In this
  paper we  present a  model that  is a generalization  of all  of the
  above  models.  It describes  models  based  on  a single  Kronecker
  product  of  spatial  and   temporal  covariance  as  well  as  more
  complicated  multi-pair models together  with any  intermediate form
  expressed  as  a sum  of  Kronecker  products  of spatial  component
  matrices of reduced rank and their corresponding temporal covariance
  matrices.   The  model  provides  a framework  for  controlling  the
  tradeoff  between the  described  complexity of  the background  and
  computational  demand for the  analysis using  this model.   Ways to
  estimate the  value of the  parameter controlling this  tradeoff are
  also discussed.
\end{abstract}

\IEEEpeerreviewmaketitle

\section{Introduction}
\label{sec:intro}
Given that  MEG/EEG is measured in  $M$ trials, on $L$  sensors and in
$C$ time samples, let ${\bf E}_m$ be the $L$ by $C$ single trial noise
matrix at  trial $m$. And to  simplify our notation  further we assume
that the  background in  ${\bf E}_m$ is  zero-mean. In this  case, the
sample spatiotemporal  covariance matrix  of dimension $N=LC$  for the
averaged over trials noise is
\begin{eqnarray}
  \mathbf{C_s} &=& \frac1{M-1} \sum_{m=1}^M 
  vec(\mathbf{E}_m)vec(\mathbf{E}_m)^T,
  \label{eq:COV}
\end{eqnarray}
where $vec(\mathbf{E})$ is all the columns of $\mathbf{E}$ stacked in
a vector.

Due  to the  tremendously  large number  of parameters  ($O(L^2C^2)$)
compared  to  the  relative  sparsity  of  background  data  typically
available  many attempts  to model  the covariance  has been  made. In
order  to be  useful for  source  analysis a  model of  spatiotemporal
covariance should capture the structure  of the background with as few
parameters  as possible.   Finally to  be useful  the model  should be
invertible, so it can be used in a likelihood formulation.

To capture the  structure in the brain background  signal we note that
there  is evidence that  the background  activity of  the brain  has a
stationary  spatial distribution  over  the time  of  interest in  the
neuroscientific experiments~\cite{RaichleME:defmbf}.  This observation
provides  for the  current approach  in exploratory  fMRI  analysis to
represent four  dimensional spatiotemporal  data as a  two dimensional
matrix  $\mathbf{X}$, which  is  decomposed into  a  sum of  Kronecker
products contaminated by normal noise~\cite{BeckmannCF:Teneic}:
\begin{equation}
  \mathbf{X} = \displaystyle\sum_{r}^{R}\mathbf{a_r}\otimes 
  \mathbf{b_r} + \mathbf{E}.
\end{equation}

Relying on such evidence, we can make assumptions about the background
activity in the brain in the MEG (magnetoencephalography)
measurements:
\begin{enumerate}
\item The measured background is a superposition of $R$ spatially
  fixed sources with independent temporal behavior.
\item Each spatial source produces correlated Gaussian noise.
\end{enumerate}

These are the basis  for modeling spatiotemporal covariance adopted in
this work. The final goal is to find spatial components according to a
justified criteria  and then estimate  respective temporal covariances
for each component. For $L$  sensors performing the measurement we can
extract at  most $L$ spatial components from  the measured background,
which  gives us the  starting form  for the  spatiotemporal covariance
matrix:
\begin{equation}
  \sum_{l = 1}^{L} \bm{\mathcal T}_{l} \otimes \bm{\mathcal S}_l,
\label{eq:general}
\end{equation}
where $\bm{{\mathcal S}}_l = v_l v_l^T$ is a spatial component
represented as a singular matrix of rank 1.

In what follows we  demonstrate how existing spatiotemporal covariance
models based on the Kronecker product can be described in the proposed
framework.  This includes the  models based  on one  Kronecker product
suggested  in~\cite{demunck2002,huizenga2002}  and   on  a  series  of
Kronecker products introduced by us~\cite{PLIS-ETAL-2004A}. Finally we
extend the framework by making it span the range of models between one
pair Kronecker models and $L$ component models.

\section{One-pair Kronecker product}
\label{sec:one_pair}
In~\cite{demunck2002} was proposed a model that casts spatiotemporal
covariance in the form that uses the Kronecker product:
\begin{eqnarray}
  \mathbf{COV} \approx \mathbf{T} \otimes \mathbf{S}.
  \label{eqn:onepair}
\end{eqnarray}
Temporal covariance is a $C\times C$ matrix {\bf T} and spatial
covariance is a $L\times L$ matrix {\bf S}, where $C$ is the number of
time samples and $L$ is the number of sensors.

An important feature of this model is its computationally fast inverse
calculation  $(\mathbf{T} \otimes  \mathbf{S})^{-1}  = \mathbf{T}^{-1}
\otimes \mathbf{S}^{-1}$.  This feature  is required of all covariance
models to make  them useful in the analysis.   All models presented in
this paper are invertible and manageable within reasonable time.

Since the  background is assumed  to have a Gaussian  distribution the
model  can  be  estimated  in  the  maximum  likelihood  framework  as
in~\cite{demunck2002}.
\begin{eqnarray}
  \mathbf{S} &=& \frac{1}{C}\left(\frac{1}{M}\displaystyle\sum_{m=1}^{M} 
    \bm{E}_m \mathbf{T}^{-1} \bm{E}_{m}^{T}\right), \\\nonumber
  \mathbf{T} &=& \frac{1}{L}\left(\frac{1}{M}\displaystyle\sum_{m=1}^{M} 
    \bm{E}_m \mathbf{S}^{-1} \bm{E}_{m}^{T}\right).
\end{eqnarray}
Here  $M$  is  the  number  of available  $L\times  C$  spatiotemporal
measurements $\bm{E}_m$.

In  essence, the  main assumption  of the  one-pair  Kronecker product
model is  that all spatial components  of the background  have one and
the same  temporal covariance structure.  To  demonstrate the validity
of  this statement we  perform spectral  decomposition of  the spatial
covariance:
\begin{equation}
\mathbf{T} \otimes \sum_{l = 1}^{L} \sigma_{l} \bm{\mathcal S}_l,
\label{eq:spectral_onepair}
\end{equation}
where  $\bm{\mathcal  S}_l  =  v_l  v_l^T$  is  an  orthonormal  basis
component  represented  as  a  singular  matrix.  Using  the  identity
$\mathbf{A}\otimes(\mathbf{B}+\mathbf{C})                             =
\mathbf{A}\otimes\mathbf{B}    +   \mathbf{A}\otimes\mathbf{C}$,   the
expression~(\ref{eq:spectral_onepair}) can be represented as:
\begin{equation}
  \sum_{l = 1}^{L} \mathbf{T} \otimes \sigma_{l} \bm{\mathcal S}_l = 
  \sum_{l = 1}^{L} \sigma_{l} \mathbf{T} \otimes \bm{\mathcal S}_{l}.
\label{eq:proto_multipair}
\end{equation}
The Left Hand Side of the equation~(\ref{eq:proto_multipair}) makes it
obvious that each spatial  component has the same temporal covariance.
The  contribution  of each  temporal  covariance  is  weighted by  the
variance of  the corresponding orthonormal spatial  component, as seen
from the  Right Hand Side of~(\ref{eq:proto_multipair}).   When we set
$\sigma_{l} \mathbf{T}  = \bm{\mathcal T}_{l}$ we see  how our general
model from~(\ref{eq:general}) subsumes the one pair model.

A parameterized  model of the same form  as in~(\ref{eqn:onepair}) was
introduced in  order to achieve a  further reduction in  the number of
parameters~\cite{huizenga2002}.  This  model is  also  a special  case
of~(\ref{eq:proto_multipair}). To see that it is enough to follow the
same logic as we just did for the unparameterized model.

\section{Multi-pair models}
\label{sec:multipair}
In  a more  realistic case  it is  easy to  picture a  situation where
several noise generators belonging to separate spatial components also
have  different  temporal  structures.   


Two     models     covering      such     case     were     introduces
in~\cite{PLIS-ETAL-2006A}    one   in    which    spatial   components
$\bm{\mathcal S}_{l}$  are components of an orthonormal  basis and the
other  with  spatial  components   from  an  independent  basis.   The
following is assumed:
\begin{enumerate}
\item Spatiotemporal  noise is  generated by $L$  spatially orthogonal
  (or  independent)  generators which  do  not  change their  location
  during the period of interest.
\item Each  spatial component is uncorrelated with other components.
\item  Measured signal  has a  Gaussian distribution  with  zero mean.
\end{enumerate}

Both  models have  the form  as in~(\ref{eq:general})  and any  set of
orthogonal  or independent  components  can be  used. As  demonstrated
in~\cite{PLIS-ETAL-2006A} estimation  of orthogonal components through
singular value decomposition and  of independent components through an
independent component analysis algorithm work well.

For given  orthogonal components $\bm{\mathcal S}_l =  v_l v_l^T$ with
$v_l$  being  a  column  of  an orthogonal  matrix  $\bm{V}$,  Maximal
Likelihood estimate of the model is:
\begin{eqnarray} \label{eq:svd_summary}
  \mathbf{COV} \approx
  \widetilde{\mathbf{COV}} &\equiv& \sum_{l=1}^L  \bm{{\mathcal T}}_l
  \otimes \bm{{\mathcal S}}_l, \\ \nonumber
  \bm{{\mathcal T}}_{l} &=& \frac 1{M} \sum_{m=1}^M 
  \mathbf{E}_{m}^{T} \bm{{\mathcal S}}_l \mathbf{E}_{m}.
\end{eqnarray}

For given  independent components $\bm{\mathcal R}_l =  w_l w_l^T$ with
$w_l$ being a row of a full rank matrix $\bm{W}^{-1}$,  Maximal
Likelihood estimate of the model is:
\begin{eqnarray} \label{eqn:ica_form}
  \mathbf{COV} \approx  
  \widetilde{\mathbf{COV}}  &\equiv& \sum_{l=1}^L \bm{{\mathcal T}}_{l}
  \otimes 
  \bm{{\mathcal R}}_{l}, \\ \nonumber
  \bm{{\mathcal T}}_{l} &=& \frac 1{M} \sum_{m=1}^M 
  \mathbf{E}_{m}^{T} 
  \mathbf{W}\mathbf{W}^{T} \bm{{\mathcal R}}_l \mathbf{W}\mathbf{W}^{T}
  \mathbf{E}_{m}.
\end{eqnarray}

These    models    are    more    general   than    the    model    of
Section~\ref{sec:one_pair} and the  independent basis multi-pair model
of~(\ref{eqn:ica_form})  is the most  general in  terms of  having the
largest  number  of degrees  of  freedom.   This increased  complexity
brings advantages in  being able to more accurately  model the complex
empirical noise covariance, which lead to improved source localization
performance  as demonstrated  in~\cite{PLIS-ETAL-2006A}.  Furthermore,
inversion is computationally efficient  due to the built in structure.
For the orthogonal basis model:
\begin{eqnarray}
  \widetilde{\mathbf{COV}}^{-1} &=& \sum_{l=1}^L 
  (\bm{{\mathcal T}}_l)^{-1} \otimes \bm{{\mathcal S}}_l.
\label{eq:ortho_inv}
\end{eqnarray}
And for the independent basis model:
\begin{equation}
  \widetilde{\mathbf{COV}}^{-1} = \sum_{l=1}^L (\bm{{\mathcal
      T}}_{l})^{-1} \otimes[ \mathbf{W} \mathbf{W}^T \bm{\mathcal
    R}_l \mathbf{W} \mathbf{W}^T].
  \label{eq:ica_inv}
\end{equation}
Details of the derivation can be found in~\cite{PLIS-ETAL-2006A}.

\section{Generalized multi-pair model}
Even  though  the  models   in  Section  \ref{sec:multipair}  are  more
descriptive  and yield  better  results than  single-pair models,  the
problem with them is the computational complexity of each iteration in
the analysis where  the likelihood needs to be  computed. The negative
log-likelihood function in  the case of using a  multipair model looks
like:
\begin{equation}
  \displaystyle\sum_{l=1}^{L}(\bm{E}_{b} - \bm{E}_{\mathcal M})^{T}
  \bm{\mathcal S}_{l} (\bm{E}_{b} - \bm{E}_{\mathcal M}) 
  \bm{\mathcal T}_{l}^{-1}.
\label{eq:cost_func}
\end{equation}
Here $\bm{E}_{b}$  and $\bm{E}_{\mathcal M}$ are  $L\times C$ matrices
of spatiotemporal measurements and  the current prediction made by the
model,   respectively.    Calculation  of~(\ref{eq:cost_func})   takes
$O(L(L^{2}C+LC^{2}+C^{3}))$ operations and it is $L$ times bigger than
for one-pair models.

Yet,  notice that  multi-pair  models utilize  all spatial  components
which in  the case of the  orthogonal basis model  also includes those
components that have small singular values. Without loss of generality
we  consider  first  only   the  orthogonal  basis  multi-pair  model.
Separating the first  $r \le L$ most significant  singular values, the
model can be expressed as
\begin{equation}
  \sum_{l = 1}^{r} \bm{\mathcal T}_{l} \otimes \bm{\mathcal S}_l +
  \sum_{k = r+1}^{L} \bm{\mathcal T}_{k} \otimes \bm{\mathcal S}_k.
\label{eq:binary_form}
\end{equation}

The  second term  in expression~(\ref{eq:binary_form})  contains small
variance  relative  to  the  first one.   Conventional  dimensionality
reduction techniques  like PCA would eliminate this  term from further
consideration. Though  this would make  the model smaller in  terms of
storage requirements it will at the same time render it useless in the
analysis.  The  inversion following the  lines of~(\ref{eq:ortho_inv})
would  become  impossible.   At  the  same time  if  we  set  temporal
covariance of all spatial components  with small singular values to be
the     same    as     it     is    assumed     in    the     one-pair
model~(\ref{eq:proto_multipair}) we obtain the following expression:
\begin{equation}
  \sum_{l = 1}^{r} \bm{\mathcal T}_{l} \otimes \bm{\mathcal S}_l +
  \bm{\mathcal T} \otimes \sum_{k = r+1}^{L} \bm{\mathcal S}_k.
\label{eq:binary_form_final}
\end{equation}

Expression~(\ref{eq:binary_form_final})  is a general  form describing
spatiotemporal  covariance models  based  on the  assumption that  the
brain has  stationary spatial distribution of  the background activity
over the time of interest~\cite{RaichleME:defmbf}. It can describe the
one-pair model when $r$ is set to zero and multi-pair models when $r =
L$.

A  feature  absolutely  required  from   this  model  is  that  it  be
invertible.   Otherwise, the model  cannot be  used in  the Likelihood
function~(\ref{eq:cost_func})  for source localization.   Indeed, this
model has  a computationally manageable inverse  with calculation time
depending on $r$ and ranging  between the times for the one-pair model
and the  multi-pair models. For the orthogonal  basis multi-pair model
the inverse is:
\begin{equation}
  \sum_{l = 1}^{r} \bm{\mathcal T}_{l}^{-1} \otimes \bm{\mathcal S}_l 
  +
  \bm{\mathcal T}^{-1} \otimes \sum_{k = r+1}^{L} \bm{\mathcal S}_k.
\label{eq:gortho_inv}
\end{equation}

To prove this claim it suffices to show that:
\begin{align} \label{eq:gortho_inv_prv}
  \left(\sum_{l = 1}^{r} \bm{\mathcal T}_{l}^{-1} 
    \otimes \bm{\mathcal S}_l +
    \bm{\mathcal T}^{-1} \otimes \sum_{k = r+1}^{L} 
    \bm{\mathcal S}_k\right) & 
  \\
  \times \left(\sum_{l = 1}^{r} \bm{\mathcal T}_{l} 
    \otimes \bm{\mathcal S}_l +
    \bm{\mathcal T} \otimes \sum_{k = r+1}^{L} 
    \bm{\mathcal S}_k\right)
  &= \bm{I}. \nonumber
\end{align}
To proceed we  need the following properties of $\bm{\mathcal S}_l$:
\begin{itemize}
\item property (1)
\begin{eqnarray}\nonumber
  (\bm{\mathcal S}_l)^2 &=& (v_l v_l^T) (v_l v_l^T) 
  = v_l (v_l^T v_l) v_l^T
  \\
  \nonumber
  &=& v_l v_l^T \quad ( v_l^T v_l = 1 \mbox{ by orthogonality})
  \\ \nonumber
  &=& \bm{\mathcal S}_l.
\end{eqnarray}
\item property (2) : $\bm{I}  = \mathbf{V} \ \mathbf{V}^T = \sum_l v_l
  v_l^T$.
\item property  (3) :  $\bm{\mathcal S}_{l} \bm{\mathcal  S}_{l^{'}} =
  \delta(l,l^{'}) \bm{\mathcal S}_{l}$, where $\delta(l,l^{'})$ is the
  Kronecker delta.
\end{itemize}
Using these  properties and performing  the multiplication in  the LHS
of~(\ref{eq:gortho_inv_prv}):
\begin{eqnarray}
  && \sum_{l = 1}^{r} \bm{I} \otimes \bm{\mathcal S}_l +
  \bm{I} \otimes \sum_{k = r+1}^{L} \bm{\mathcal S}_k \\
  &=&  \bm{I} \otimes \sum_{l = 1}^{r} \bm{\mathcal S}_l +
  \bm{I} \otimes \sum_{k = r+1}^{L} \bm{\mathcal S}_k \nonumber\\
  &=&  \bm{I} \otimes \sum_{l = 1}^{L} \bm{\mathcal S}_l \nonumber\\
  && \mbox{and using property (2)}\nonumber\\
  &=& \bm{I} \otimes \bm{I} = \bm{I}. \nonumber
\end{eqnarray}

For the  independent basis multi-pair  model if we  replace orthogonal
components $\bm{\mathcal S}_l$ by independent components $\bm{\mathcal
  R}_l$ the inverse is:
\begin{eqnarray} \label{eq:gica_inv}
  \sum_{l = 1}^{r} \bm{\mathcal T}_{l}^{-1} \otimes \bm{WW}^T 
  \bm{\mathcal R}_l \bm{WW}^T \\ +
  \bm{\mathcal T}^{-1} \otimes \sum_{k = r+1}^{L} 
  \bm{WW}^T \bm{\mathcal R}_k \bm{WW}^T. \nonumber
\end{eqnarray}
The proof proceeds in the  similar fashion as for the orthogonal basis
model.

Computational  efficiency  of   this  new  generalized  spatiotemporal
covariance model is now a function  of $r$ and not fixed to the number
of sensors, as in the case of models from Section~\ref{sec:multipair}.
Here again,  without loss of  generality we demonstrate  results using
the   orthogonal   basis   model.    We   redefine   $\sum_{k=r+1}^{L}
\bm{\mathcal S}_{k}  = \bm{\mathcal S}$.   The negative log-likelihood
function with the use of generalized covariance is then expressed as:
\begin{eqnarray} \label{eq:cost_func_gen}
  &&\displaystyle\sum_{l=1}^{r}(\bm{E}_{b} - \bm{E}_{\mathcal M})^{T}
  \bm{\mathcal S}_{l} (\bm{E}_{b} - \bm{E}_{\mathcal M}) 
  \bm{\mathcal T}_{l}^{-1} \\ \nonumber
  &+&(\bm{E}_{b} - \bm{E}_{\mathcal M})^{T} \bm{\mathcal S} 
  (\bm{E}_{b} - \bm{E}_{\mathcal M}) \bm{\mathcal T}^{-1}.
\end{eqnarray}
The    computational    complexity     of    this    calculation    is
$O((r+1)(L^{2}C+LC^{2}+C^{3}))$   and   is   flexible  since   it   is
parameterized   by   $r$.   It   can   be   as   large  as   the   one
of~(\ref{eq:cost_func}) when  the background is  diverse enough to
need  a complex  representation and  can be  as small  as that  of the
one-pair model.

\addtolength{\textheight}{-4.4cm}

The         generalized         orthogonal         basis         model
from~(\ref{eq:binary_form_final})  can be  estimated  in a  relatively
straightforward  manner based  on the  nature  of PCA.  The number  of
single  Kronecker product  pairs of  our  model $r+1$  is obtained  by
choosing the  most significant  components provided by  singular value
decomposition.  This is  a well justified approach and  it follows the
conventional method of dimensionality reduction in PCA. After that the
model can  be estimated in  the Maximal Likelihood  framework.  Before
proceeding with  its derivation we  need to calculate  the determinant
of~(\ref{eq:binary_form_final}).  It can be found from the determinant
of     the     orthogonal     basis    multipair     model     derived
in~\cite{PLIS-ETAL-2006A}:
\begin{equation}
  |\displaystyle\sum  \bm{\mathcal T}_{l} 
    \otimes \bm{\mathcal S}_{l}| = 
  \displaystyle \prod_{l}^{L} | \bm{\mathcal T}_{l}|.
  \label{eq:svd_det}
\end{equation}
Noticing that the last $J=L-r$ temporal covariances in the product are
the same, and are $\bm{\mathcal T}$, we get:
\begin{equation}
  |\sum_{l = 1}^{r} \bm{\mathcal T}_{l} \otimes \bm{\mathcal S}_l +
  \bm{\mathcal T} \otimes \bm{\mathcal S}| =
  \bm{\mathcal T}^{J} \displaystyle \prod_{l}^{r} | \bm{\mathcal T}_{l}|.  
\label{eq:det_gen}
\end{equation}
The log-likelihood function ${\mathcal L} $ is then written as:
\begin{align}
   &const-\frac{M}{2}\left(J\ln| \bm{\mathcal T}| + 
    \sum_{l=1}^{r} \ln|\bm{\mathcal T}_{l}|\right) \\ \nonumber 
   &- \frac{1}{2}tr\left(\sum_{m=1}^{M} \left( \sum_{l=1}^{r} 
      \mathbf{E}_{m}^{T}  \bm{{\mathcal S}}_l \mathbf{E}_{m} 
      (\bm{\mathcal T}_{l})^{-1} +   
      \mathbf{E}_{m}^{T}  \bm{\mathcal S} \mathbf{E}_{m} 
      \bm{\mathcal T}^{-1} \right) \right).
\end{align}
Differentiating with respect to $\bm{\mathcal T}$ results in:
\begin{align} 
d{\mathcal L} = & -\frac{MJ}{2}tr((\bm{\mathcal T})^{-1} 
d\bm{\mathcal T}) \\
& + 
\frac{1}{2}tr\left(\sum_{m=1}^{M}\mathbf{E}_{m}^{T} \bm{\mathcal S}
\mathbf{E}_{m} 
(\bm{\mathcal T})^{-1} d\bm{\mathcal T} 
(\bm{\mathcal T})^{-1}\right) 
\nonumber \\
=& -\frac{MJ}{2}tr((\bm{\mathcal T})^{-1}d\bm{\mathcal T}) \nonumber \\ 
& + 
\frac{1}{2}tr\left(\sum_{m=1}^{M} (\bm{\mathcal T})^{-1} 
\mathbf{E}_{m}^{T} 
\bm{\mathcal S} \mathbf{E}_{m}
(\bm{\mathcal T})^{-1} d\bm{\mathcal T}\right) \nonumber \\
=& -\frac{MJ}{2} \nonumber \\
&tr\left(\left[(\bm{\mathcal T}^{l})^{-1} \nonumber  
-\frac{1}{M}\sum_{m=1}^{M}(\bm{\mathcal T})^{-1} \mathbf{E}_{m}^{T} 
\bm{\mathcal S} \mathbf{E}_{m}
(\bm{\mathcal T})^{-1} \right] d\bm{\mathcal T} \right) \nonumber
\end{align}
And the final result is:
\begin{equation}
  \bm{\mathcal T} = \frac{1}{MJ}\sum_{m=1}^{M}\mathbf{E}_{m}^{T} 
  \bm{\mathcal S} \mathbf{E}_{m}.
  \label{eq:mle_svd_gen}
\end{equation}
Together with the  equation~(\ref{eq:svd_summary}) we can estimate the
generalized  model as  a sum  of $r$  Kronecker products  of  rank one
spatial   component  matrices   with   their  corresponding   temporal
covariance matrices  plus a Kronecker  product of a  spatial component
matrix of rank $J$ with its corresponding temporal covariance matrix.

This approach  will not work  for the independent basis  model.  There
are no singular values that can be used for thresholding.  However, we
can  use the  variance of  the  data projected  into each  independent
spatial component to perform thresholding  based on its value.  Then a
similar Maximal  Likelihood estimation as  in~(\ref{eqn:ica_form}) can
be used for  the generalized independent basis model.   This would add
asymmetry to the way these two models are estimated.

Another  possible   approach  can   be  based  on   observations  made
in~\cite{PLIS-ETAL-2004A}.   Many  temporal  covariances of  different
spatial  components were  found  to be  similar.   By discovering  $r$
clusters of  similar temporal covariances we can  obtain $r$ Kronecker
products  of  the  generalized  multi-pair model.   This  approach  is
applicable  both  for  orthogonal  and independent  basis  models.   A
similarity criterion  could be developed  for that case, such  that it
provides clustering of components in a manner that balances optimizing
source localization with reducing  computational time.  In this second
approach, the resulting model will  be different from the one obtained
by thresholding of significant components. The model is then described
as  a sum  of  Kronecker  products of  spatial  component matrices  of
incomplete  rank and  their corresponding  temporal  covariances.  The
difference  is  that  there  is  no requirement  of  having  rank  one
component matrices and  having only one spatial component  matrix of a
bigger rank as in~(\ref{eq:binary_form_final}).

It  is to be  determined which  of these  two approaches  shows better
localization  performance.  In future  work  we  plan  to answer  this
question  and  investigate  which  clustering  algorithm  to  use  for
obtaining summands in the second approach.

\bibliographystyle{plain}
\bibliography{../../covariance_models/Covariance-Neuroimage}

\end{document}